\documentclass[conference]{IEEEtran}
\IEEEoverridecommandlockouts
\usepackage{fancyhdr}
\usepackage{graphicx}
\usepackage{amsmath}
\usepackage{amssymb}
\usepackage{wrapfig}
\usepackage{soul}
\usepackage{xcolor}
\usepackage{verbatim}
\usepackage{subcaption}
\usepackage{tikz}
\usetikzlibrary{spy}
\usepackage{algorithm}
\usepackage{textcomp}
\usepackage[export]{adjustbox}
\usepackage{siunitx}
\usepackage{array}
\usepackage{commath}
\usepackage{url}
\usepackage{paralist}
\usepackage{booktabs} 
\usepackage{algpseudocode}
\usepackage{gensymb}
\usepackage{enumitem}
\usepackage{cancel}
\setlist{nolistsep}
\graphicspath{ {./figs/} }

\def\BibTeX{{\rm B\kern-.05em{\sc i\kern-.025em b}\kern-.08em
    T\kern-.1667em\lower.7ex\hbox{E}\kern-.125emX}}
    
\newcommand{\argmin}{\arg\!\min}

\algnewcommand\algorithmicinput{\textbf{INPUT:}}
\algnewcommand\INPUT{\item[\algorithmicinput]}
\algnewcommand\algorithmicoutput{\textbf{OUTPUT:}}
\algnewcommand\OUTPUT{\item[\algorithmicoutput]}
\algnewcommand\algorithmiclocal{\textbf{LOCAL:}}
\algnewcommand\LOCAL{\item[\algorithmiclocal]}
\algnewcommand\algorithmicparam{\textbf{PARAMETERS:}}
\algnewcommand\PARAM{\item[\algorithmicparam]}
\algnewcommand\algorithmicinout{\textbf{INOUT:}}
\algnewcommand\INOUT{\item[\algorithmicinout]}

\newcommand{\pluseq}{\mathrel{+}=}

\begin{document}

\title{Image Gradient Decomposition for Parallel and Memory-Efficient Ptychographic Reconstruction\\
\thanks{
This manuscript has been authored by UT-Battelle, LLC, under contract DE-AC05-00OR22725 with the US Department of Energy (DOE). The publisher acknowledges the US government license to provide public access under the DOE Public Access Plan (http://energy.gov/downloads/doe-public-access-plan).
This research is sponsored by the AI Initiative as part of the Laboratory Directed Research and Development Program of Oak Ridge National Laboratory, managed by UT-Battelle, LLC, for the US Department of Energy under contract DE-AC05-00OR22725. This research used resources at the Oak Ridge Leadership Computing Facility, a DOE Office of Science User Facility operated by the Oak Ridge National Laboratory.
}}

\author{
\IEEEauthorblockN{Xiao Wang\IEEEauthorrefmark{1}, 
Aristeidis Tsaris\IEEEauthorrefmark{1}, Debangshu Mukherjee\IEEEauthorrefmark{1}, Mohamed Wahib\IEEEauthorrefmark{2}, \\
Peng Chen\IEEEauthorrefmark{3},  
Mark Oxley\IEEEauthorrefmark{1},
Olga Ovchinnikova\IEEEauthorrefmark{1},
Jacob Hinkle\IEEEauthorrefmark{1}}
\IEEEauthorblockA{\IEEEauthorrefmark{1}Oak Ridge National Laboratory, Oak Ridge, United States
\\\{wangx2, tsarisa, mukherjeed, oxleymp, ovchinnikovo,hinklejd\}@ornl.gov}
\IEEEauthorblockA{\IEEEauthorrefmark{2}RIKEN Center for Computational Science, Tokyo, Japan
\\mohamed.attia@riken.jp}
\IEEEauthorblockA{\IEEEauthorrefmark{3}National Institute of Advanced Industrial Science and Technology, Tokyo, Japan
\\chin.hou@aist.go.jp}
}


\maketitle
\thispagestyle{fancy}
\lhead{}
\rhead{}
\chead{}
\lfoot{\footnotesize{
SC22, November 13-18, 2022, Dallas, Texas, USA
\newline 978-1-6654-5444-5/22/\$31.00 \copyright 2022 IEEE}}
\rfoot{}
\cfoot{}
\renewcommand{\headrulewidth}{0pt}
\renewcommand{\footrulewidth}{0pt}

\begin{abstract}
Ptychography is a popular microscopic imaging modality for many scientific discoveries and sets the record for highest image resolution. Unfortunately, the high image resolution for ptychographic reconstruction requires significant amount of memory and computations, forcing many applications to compromise their image resolution in exchange for a smaller memory footprint and a shorter reconstruction time. In this paper, we propose a novel image gradient decomposition method that significantly reduces the memory footprint for ptychographic reconstruction by tessellating image gradients and diffraction measurements into tiles. In addition, we propose a parallel image gradient decomposition method that enables asynchronous point-to-point communications and parallel pipelining with minimal overhead on a large number of GPUs. Our experiments on a Titanate material dataset (PbTiO\textsubscript{3}) with 16632 probe locations show that our Gradient Decomposition algorithm reduces memory footprint by 51 times. In addition, it achieves time-to-solution within 2.2 minutes by scaling to 4158 GPUs with a super-linear strong scaling efficiency at 364\% compared to runtimes at 6 GPUs. This performance is 2.7 times more memory efficient, 9 times more scalable and 86 times faster than the state-of-the-art algorithm.
\end{abstract}


\section{Introduction}
\label{sec:prob_overview} 

Ptychography is a popular imaging modality for many scientific discoveries and reconstructs an object from a set of spatially overlapped X-ray or electron diffraction measurements~\cite{Zheng13}. With diffraction technology, ptychography removes the theoretical resolution limit determined by microscope lenses and allows microscopic image resolution to be even higher. Therefore, ptychographic imaging sets the current records for the highest microscope resolution and many applications, such as material imaging~\cite{Bhartiya21}, human biology imaging~\cite{Chen20} and security encryption imaging~\cite{Shi13}, rely on its high resolution to observe features for small particles~\cite{Kahnt21, Jiang18}.

Despite its capability in delivering unprecedented image resolution, ptychographic imaging requires enormous memory to store diffraction measurements and 3D image pixels, also called ``{\em Voxels}"~\cite{yu2021,Nashed14,Yu2021-ics}. This memory requirement, in turn, constrains the achievable ptychographic image resolution and stops many scientists from discovering important features that are present in the samples. Furthermore, ptychographic imaging often requires real-time reconstruction while collecting diffraction measurements and use the reconstruction to guide the data acquisition on-the-fly. Unfortunately, ptychographic reconstruction is too slow and often fails to guide experiments in time due to the memory constraints and the significant number of computations.  

To address the memory constraints and improve reconstruction speed, the state-of-the-art Halo Voxel Exchange algorithm~\cite{yu2021,Nashed14,Yu2021-ics} takes advantage of the fact that ptychography diffraction measurements are often organized into a sequence of matrices indexed in time order. Each time step of the data acquisition is called a ``{\em Probe Location}" and the diffraction measurements at a probe location corresponds to a group of spatially contiguous voxels in the shape of a circle in the image domain. Neighboring probe locations correspond to different but overlapped circles in the image. To reduce memory, the Halo Voxel Exchange algorithm uses a grid pattern to distribute the reconstruction volume and diffraction measurements among GPUs. To avoid potential conflicts and synchronize neighboring tiles at their borders, each tile's borders are assigned with additional voxels and measurements from the neighboring tiles~\cite{Nashed14}. Then, each tile is updated in parallel and the updated border voxels are exchanged among neighboring tiles through remote memory copy operations.

The Halo Voxel Exchange method, however, has several issues. (1) Its memory reduction capability is limited. The number of additional border voxels and measurements assigned to each GPU takes significant amount of memory, causing a memory capacity bottleneck and poor memory reduction capability.
(2) It has a poor scalability performance because of the halo exchange operations and the redundant computations from the additional border voxels and diffraction measurements. (3) Pasting border voxels to neighboring halos creates unacceptable artificial image seam border artifacts and worsens image quality~\cite{Nashed14}.

In response to the issues for the Halo Voxel Exchange method, this paper presents an image gradient decomposition method for 3D ptychographic reconstruction. It uses image gradients and accumulated image gradients buffers, instead of voxels, to synchronize reconstructions across different tiles, without the need of assigning any additional probe locations to any tile. Therefore, the memory reduction capability performs much better than the Halo Voxel Exchange method. To achieve a good scalability with minimal parallel overhead, The Gradient Decomposition method uses asynchronous point-to-point communications as well as parallel pipelining to exchange image gradients and gradients buffers among GPUs while hiding its communication cost. Furthermore, the Gradient Decomposition method can completely eliminate the artificial image border artifacts that are present in the Halo Voxel Exchange method.

In summary, this paper makes three contributions: (1) proposing a novel gradient based decomposition method for ptychographic reconstruction and reducing memory footprint, without introducing any redundant computations or image seam artifacts like the Halo Voxel Exchange method; (2) 
introducing parallel forward and backward gradient accumulation operations to synchronize gradients among image tiles; and (3) presenting an asynchronous pipelining framework to reduce communication cost across GPUs. Our experimental results on a large-scale Lead Titanate dataset with 16632 probe locations demonstrate a 2519 times speedup for the proposed Gradient Decomposition algorithm with a super-linear strong scaling efficiency at 364\% by scaling from 6 GPUs to 4158 GPUs. At the same time, the average GPU memory footprint reduces by 51 times. This performance is 2.7 times more memory efficient, 9 times more scalable and 86 times faster than the state-of-the-art Halo Voxel Exchange algorithm.

  \section{Background}
\label{sec:background}
\begin{figure}
   \includegraphics[width=\linewidth]{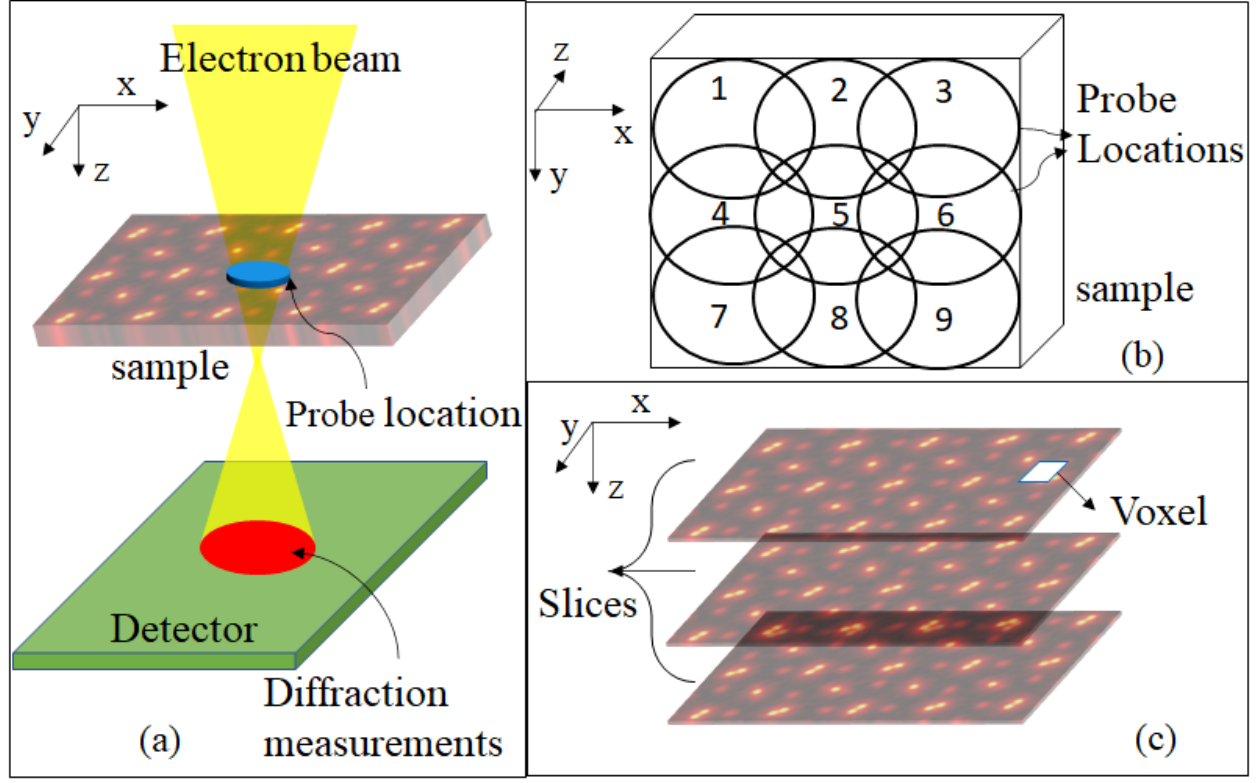}
 \caption{(a) Shows a setup for electron ptychography. (b) Demonstrates the raster order scan pattern with each circle representing a probe location. Each probe location circle is also indexed with a number to indicate its corresponding time order for data acquisition. In addition, notice that the probe location circles overlap with each other. (c) Illustrates the concept of image slices and voxels.
   \label{fig:ptycho_setup} 
    } 
\end{figure}

\subsection{Ptychography Setup}
\label{sec:ptycho_setup} 
At each probe location, an electron microscope probe generates electron beams to penetrate a material sample, and the electron beams incident on the sample in the shape of a circle, exemplified in Fig.~\ref{fig:ptycho_setup}(a). After the electron beams penetrate the sample, the exit electron beams are diffracted and a detector captures the diffraction measurements on the other end of the material sample across from the electron probe. In the example of Fig.~\ref{fig:ptycho_setup}(a), we use a blue circle to represent a probe location circle and we use a red circle to represent the diffraction measurements taken by the detector at that probe location. In addition, Fig.~\ref{fig:ptycho_setup}(a) uses a coordinate system where z points downward along the direction of the electron beams and the detector is perpendicular to z and is in the x-y plane.

To generate enough diffraction measurements for reconstruction, the electron probe is not stationary. Instead, it scans across the sample in a raster order in small steps and diffraction measurements are recorded by the detector at each probe location. Fig.~\ref{fig:ptycho_setup}(b) shows an example scan pattern (we rotate the coordinate system for an easier view for readers). The y axis now points downward, the x axis points to the right, and the z axis points away from the x-y plane. Each circle represents a probe location circle and the number in the circle represents the data acquisition time order for the respective probe location. In this example, the electron microscope starts the data acquisition at probe location 1, and moves its probe from location 1 to 2, 3 … until it reaches location 9. Also, it is important to mention that the neighboring probe location circles overlap with each other to ensure a good image spatial resolution without artifacts. In fact for most ptychographic imaging applications, the probe location circles overlap ratio is larger than 70\% to ensure a good image quality~\cite{Huang17}.

\begin{figure*}
\centering
\includegraphics[width=\linewidth]{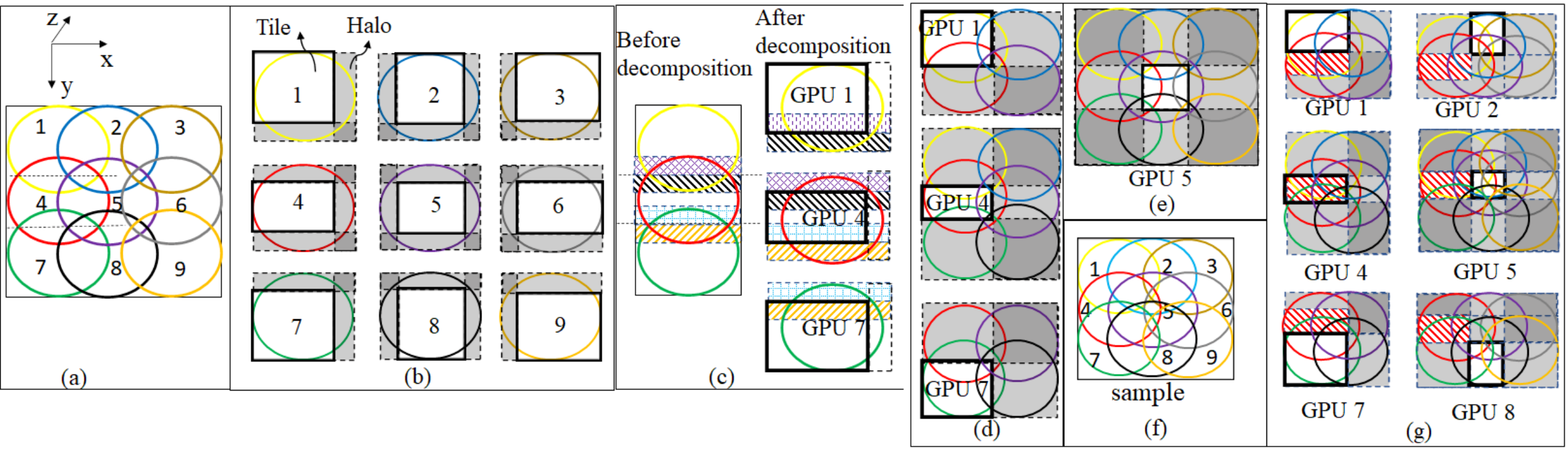}
 \caption{(a) A 9 probe locations scan pattern with a different color for each circle. (b) Each GPU is assigned a tile and a probe location circle. Each tile is also extended with halos to cover the entire probe location circle. (c) The same color shaded regions conflict with each other because each tile only has partial measurements. (d) An example of embarrassingly parallel updates for a dataset with a low probe location overlap ratio. Each tile is assigned with additional probe locations and the halos are augmented. (e) Notice that the center tile at GPU 5 has to be assigned with all probe locations. (f) An example with a large probe location overlap ratio. (g) Shows the halo voxel exchange arrangement. The tile at GPU 4 is shaded in red, and is copy-pasted to the same color halos at all neighboring tiles.} 
   \label{fig:halo_voxel_exch1} 
 
\end{figure*}

\subsection{Related Work and Math Formulation}
\label{sec:math_formulation} 
There are two classes of algorithms for electron ptychographic reconstruction, Fourier Deconvolution and Maximum Likelihood. The Fourier Deconvolution reconstruction methods, such as Single-Shot Fourier Ptychography~\cite{Sidorenko16} and Wigner Distribution Deconvolution~\cite{Li2014}, relies on Fourier Transform and inverse Fourier Transform to reconstruct images from measurements. Although the Fourier Deconvolution methods are easy to compute and highly parallel, these methods have four major disadvantages. (1) The Fourier Deconvolution reconstruction methods are sensitive to noise in the measurements and thereby require high electron radiation doses for data acquisition to minimize noise; (2) They have poor performance for certain dose reduction techniques, such as defocused electron microscopy~\cite{Jiang18};  (3) The Fourier Deconvolution methods produce inferior image quality in the presence of imaging system aberration and defects;  (4) They can only reconstruct 2D images, and cannot reconstruct 3D volumes.

In contrast, the Maximum Likelihood reconstruction methods or its variants, such as the Ptychographical Iterative Engine (PIE)~\cite{Maiden17}, are 3D reconstruction methods that iteratively compute a Maximum Likelihood cost function while addressing all four issues for the Fourier Deconvolution methods. They can (1) reduce the needed radiation doses and are compatible for defocused data acquisition; (2) are less sensitive to measurements noise while producing superior image quality compared to Fourier Deconvolution~\cite{Jiang18}; (3) and correct microscope aberration and defects in the reconstruction through complex imaging system modeling. Unfortunately, the system modeling in the Maximum Likelihood methods significantly increases the amount of computations and make parallelizing the computations difficult. A detailed description for the parallel computation challenges for the Maximum Likelihood methods are described later in Sec.~\ref{sec:halo-voxel-exchange}. 

Mathematically, the Maximum Likelihood methods can be described as the solution to the following non-linear cost function, $F(V)$:
\begin{equation}
    V \gets \argmin_V F(V) = \argmin_V \sum_{i=1}^{N}  \left( | y_i | - | G(p_i,V) |   \right)^2    \ , 
\label{eqn:math_formulation}
\end{equation}
In the above equation, $V$ is the result of 3D ptychographic reconstruction and represents the scattering potential for the sample to be imaged and each element of $V$ is a 3D pixel, also known as ``voxel". $V$ can also be viewed as a stack of 2D image slices, and each image slice is a cross section of the 3D reconstruction with a fixed thickness. Fig.~\ref{fig:ptycho_setup}(c) shows an example of a 3D image reconstruction where the sample consists of 3 image slices
Each image slice is perpendicular to the z axis and an example voxel is shown as a white square in the figure. In the above equation, $i$ represents the probe location index, ranging from the first probe location to the $N^{\it th}$ probe location. $y_i$ represents the Fourier transform for diffraction measurements from the $i^{\it th}$ probe location, and $| y_i|$ represents the magnitude of $y_i$.
$p_i$ is a probe function that represents the physics properties of the microscope at the $i^{\it th}$ probe location.

The function $G$ in the above equation is a computationally expensive and non-linear multi-slice function that can simulate diffraction measurements at the $i^{\it th}$ probe location from probe function $p_i$ and image reconstruction $V$~\cite{Maiden12}. To do that, $G$ computes a Fourier transform and an inverse Fourier transform for each image slice in sequence, and the output from the previous image slice is used as the input for the next slice until reaching the last slice.
After obtaining the simulated measurements through function $G$, Eqn.~(\ref{eqn:math_formulation}) computes the magnitude of the Fourier transform for the simulated measurements, $|G(p_i,V)|$, and compares it with the magnitude of the ground truth measurements acquired through data acquisition, $| y_i|$, at every probe location to determine how realistic the image reconstruction $V$ is in matching the simulated with the ground truth.
If the difference between the simulated and the ground truth measurements is large, reconstruction $V$ is updated using gradient descent methods until the difference of the two is small.

\begin{figure*}
\centering
\includegraphics[width=\linewidth]{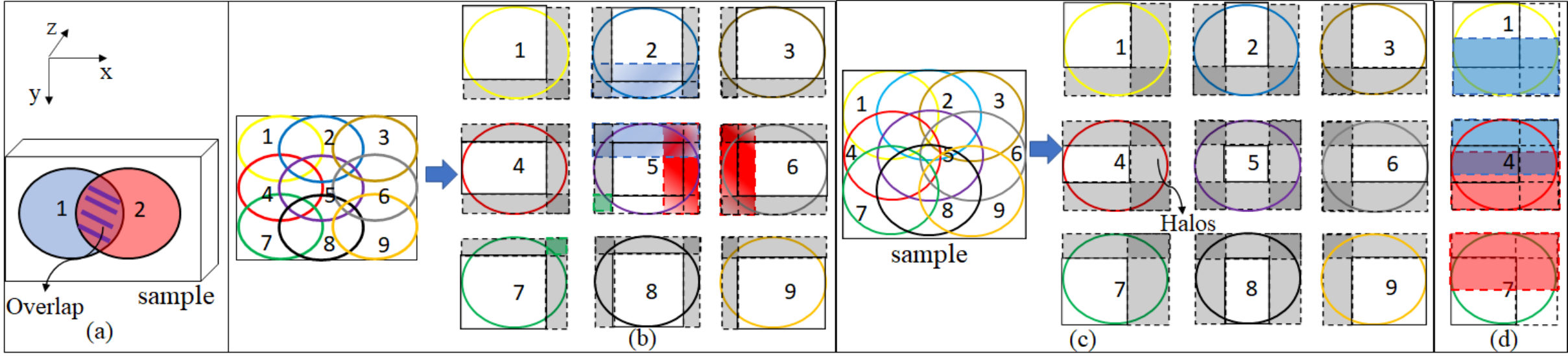}
 \caption{(a) Two overlap probe location circles with an individual image gradient for each. The individual image gradients in the purple shaded overlap region are added together. (b) Shows the decomposed image gradient tiles with extended halos for the 9 circle example in Fig.~\ref{fig:halo_voxel_exch1}(a). Notice that the halos are much smaller than those for the Halo Voxel Exchange method in Figs.~\ref{fig:halo_voxel_exch1}(d) and (e). (c) Shows the image gradient tiles for the high overlap ratio example in Fig.~\ref{fig:halo_voxel_exch1}(f). (d) Tile 1 overlaps with both tiles 4 and 7. However, the image gradients at tile 1 are added to tile 4, but not tile 7.   
   \label{fig:gradient} 
    } 
\end{figure*}

\subsection{Halo Voxel Exchange}
\label{sec:halo-voxel-exchange} 
The state-of-the-art parallel algorithm for the Maximum Likelihood reconstruction methods are the Halo Voxel Exchange method~\cite{Nashed14,Yu2021-ics,yu2021}. To explain how it works, we reuse the 9 overlap probe locations example from Fig.~\ref{fig:ptycho_setup}(b) in Fig.~\ref{fig:halo_voxel_exch1}(a), but with a different color for each circle. If we assume that there are 9 GPUs in total, the Halo Voxel Exchange method achieves memory footprint reduction not only by distributing the image into 9 contiguous tiles among the 9 GPUs, but also by distributing the probe location measurements among the GPUs in the same way. In the example of Fig.~\ref{fig:halo_voxel_exch1}(b), GPU 1 is assigned with the measurements for the yellow probe location circle and an image tile highlighted in black. GPU 2 receives the blue probe location circle and a different tile highlighted in black.

An issue we can notice from Fig.~\ref{fig:halo_voxel_exch1}(b) is that each tile does not entirely cover its probe location circles and some circles are partially outside the tiles. Therefore, the Halo Voxel Exchange method extends each tile with halos, shown as gray shaded rectangles in Fig.~\ref{fig:halo_voxel_exch1}(b), to cover the probe location circles.
The goal is then to let each GPU perform parallel reconstruction independently for their extended tiles until the algorithm converges. Then, the GPUs abandon their halos but stitch together their tiles, and the stitched result should be the same as the ground truth reconstruction in Fig.~\ref{fig:halo_voxel_exch1}(a)~\cite{Nashed14}.
Unfortunately, the independent tile reconstructions conflict with each other. For example, the black shaded region in Fig.~\ref{fig:halo_voxel_exch1}(c) have two independent copies after the decomposition. One copy is in the black shaded halo for GPU 1, and is reconstructed with the measurements from the yellow probe location circle alone. The other copy is in the same color tile for GPU 4, but is reconstructed with measurements from the red circle alone. Neither of them, however, has all the needed measurements and neither has the correct reconstruction. Similarly, the purple shaded tile for GPU 1 has a conflict with the same color halo for GPU 4 and both of them only have partial measurements.

To address the above issue and independently reconstruct tiles in parallel, the Halo Voxel Exchange method assigns additional probe locations that are neighbors to the probe location at each tile. For example, Fig.~\ref{fig:halo_voxel_exch1}(d) has a low probe locations overlap ratio and GPU 1 is not only assigned with the yellow circle, but also has the neighboring red, blue and purple circles. The halo at GPU 1 is augmented accordingly to cover all the additional circles. Similarly, GPU 4 has five additional probe locations from the neighboring yellow, blue, purple, green, and black circles and the halos are augmented. With the additional probe locations, each GPU has all the needed overlap probe locations and can perform independent tile reconstructions in parallel without any communication until the algorithm converges. For the final result, the GPUs simply abandon their halos but stitch together their tiles.   

The additional neighboring probe locations and the embarrassingly parallel updates, however, have a couple issues. First, the additional probe locations significantly diminish the memory reduction capability. Before assigning the additional probe locations, each GPU only has 1 probe location. After assigning the additional locations, GPU 1 has 4 probe locations and GPU 4 has 6 probe locations. In the extreme case of GPU 5 at the center tile in Fig.~\ref{fig:halo_voxel_exch1}(e), it has all 9 probe locations, thereby achieving no memory reduction at all.

The second issue is that the embarrassingly parallel updates are useful only when the probe overlap ratio is low.  Ptychographic imaging often has a high overlap ratio among probe location circles, sometimes more than 80\%. When the overlap ratio is high enough that a probe location circle not only overlaps with its adjacent neighbors, but also overlaps with non-adjacent neighbors, such as in Fig.~\ref{fig:halo_voxel_exch1}(f), the embarrassingly parallel updates fail to produce correct reconstructions. For example, the red shaded tile for GPU 4 in Fig.~\ref{fig:halo_voxel_exch1}(g) and the same color halos for GPUs 1, 2, 5, 7 and 8 are all different reconstructions for the same region overlapped by many probe locations. The red shaded tile for GPU 4 has all the measurements from 6 different location circles. The red shaded halo for GPU 1, however, does not have the measurements from the green and black probe locations and therefore is inconsistent with the tile for GPU 4. Similarly, GPUs 2, 5, 7 and 8 from the neighboring tiles do not have all the measurements for their red shaded halos.

To tackle the above problem for datasets with a high probe overlap ratio, the Halo Voxel Exchange method assigns additional probe locations and independently reconstructs the tiles in parallel as before in Figs.~\ref{fig:halo_voxel_exch1} (d) and (e). At the end of the independent tile reconstruction, the voxels in each tile are pasted to the halos in neighboring GPUs through synchronous point-to-point communications. For example, the red tile from GPU 4 is pasted to the same color halos in neighboring GPUs 1, 2, 5, 7 and 8. After the copy-pastes, the tiles are now consistent with each other. Then, the independent tile reconstructions and voxels copy-paste operations are repeated until the algorithm converges.

The Halo Voxel Exchange method, however, has a few fundamental limitations. The first issue is its poor memory reduction capability because of the additional memory allocation for the augmented halos and the extra probe location measurements, as explained before in Figs.~\ref{fig:halo_voxel_exch1}(d) and (e). The second issue is its poor scalability because the extra probe location measurements lead to many redundant computations and the overhead to exchange tiles and halos among GPUs is large. The third issue is that the voxel copy-paste operations cause image seam border artifacts at the border of every tile~\cite{Nashed14}. In this paper, we demonstrate a gradient-based image decomposition method that addresses all these issues collectively. Experiments on two Lead Titanate material datasets with 4158 and 16632 probe locations show that our Gradient Decomposition algorithm eliminates the image border artifacts, is 86 times faster, 9 times more scalable and 2.7 times more memory efficient than the Halo Voxel Exchange on 4158 GPUs.
\section{Image Gradient Decomposition}
\label{sec:image_gradient}

An image gradient is defined as the derivative of the cost function $F(V)$ with respect to reconstruction $V$, and is denoted as $\frac{\partial F}{\partial V}$. If we use $f_i(V)$ to represent the square error between the actual and the simulated diffraction measurements at probe location $i$, namely $f_i(V)=\left( \| y_i\| - \|G(p_i,V)\|   \right)^2$, then the total image gradient, $\frac{\partial F}{\partial V}$, can be expressed as the summation of individual image gradients $f_i(V)$ across all probe locations:
\begin{equation}
\frac{\partial F}{\partial V} = \frac{\partial \sum_{i=1}^{i=N}  f_i(V)}{\partial V} =  \sum_{i=1}^{i=N} \frac{\partial f_i(V)}{\partial V}
\label{eqn:summation_gradient}
\end{equation}

A special property to simplify the above computations is that the magnitude for the individual image gradient, $\frac{\partial f_i(V)}{\partial V}$, is significant only within the probe location circle $i$,  and is almost zeros everywhere outside its circle because $f_i(V)$ is constant outside the circle. For example, Fig.~\ref{fig:gradient}(a) has two probe locations. The individual image gradient for probe location 1,$\frac{\partial f_1(V)}{\partial V}$, is significant within the probe location circle 1. Outside of its circle, the individual image gradient is almost zeros everywhere and has no impact on image reconstruction. The same can be said for the individual image gradient for probe location 2. To apply this special property, we use Fig.~\ref{fig:gradient}(a) again for illustration and use 2 GPUs with each assigned with a probe location. Then each GPU can independently compute their individual image gradients for their probe location, but adds their individual image gradients together for the overlap purple shaded region in Fig.~\ref{fig:gradient}(a). The total image gradient, $\frac{\partial F}{\partial V}$, is then the concatenation of the individual image gradients from the non-overlap blue and red regions and the accumulated image gradients from the purple shaded overlap region.

With the above special property, we reuse the example in Fig.~\ref{fig:halo_voxel_exch1}(a) with 9 overlap probe location circles to demonstrate how the Gradient Decomposition method works. If we have 9 GPUs, the Gradient Decomposition method divides the image gradients and probe locations into 9 tiles among the GPUs, and the tiles are extended with gray halos in the figure to cover their probe location circles.

To accumulate image gradients correctly for the overlap regions, the gradients at each tile's overlap region need to be added to the gradients at the same overlap region in all the neighboring tiles. We use the center tile 5 in Fig.~\ref{fig:gradient}(b) for example. Tile 5 overlaps with tile 2 in the blue overlap region, which includes both a halo and a partial tile. Then the GPU at tile 5 exchanges its image gradients at the blue overlap region with the GPU at tile 2, by \textbf{adding} their image gradients to each other for the blue overlap region through point-to-point communications. Similarly, tile 5 also overlaps with tile 6 in the red overlap region, and the GPUs at tile 5 and tile 6 \textbf{add} to each other’s image gradients at the red overlap region. 
Then, the same gradient exchange and accumulation operations are applied to every tile and its direct neighboring tiles, including the neighboring tiles in the diagonal direction. For example, tiles 1, 3, 7, 9 overlap with center tile 5 in their corners and the overlaps take the shape of small squares, such as the green squares at the corner of tiles 7 and 5.
After the image gradients at the overlap regions are accumulated, each GPU uses a gradient descent algorithm to independently update their tiles based on its image gradients. If the algorithm does not converge, the Gradient Decomposition method repeats itself and image gradients are recomputed until convergence is achieved. The final image reconstruction is the result of stitching together the non-halo tiles after throwing away the halos.

It is important to note that there are two major differences between the Gradient Decomposition and the Halo Voxel Exchange methods. First, the Gradient Decomposition does not assign any extra probe locations to each tile as in Halo Voxel Exchange. Therefore, if we compare the halo sizes in Fig.~\ref{fig:gradient}(b) with those in Figs.~\ref{fig:halo_voxel_exch1}(d) and (e), we notice that the halo sizes for Gradient Decomposition are much smaller than those for Halo Voxel Exchange method, leading to a smaller memory footprint and fewer computations. Our experiment results for a large-scale Lead Titanate material dataset with 16632 probe locations show in Table~\ref{table:performance-large} that the Gradient Decomposition consistently has a smaller memory footprint than the Halo Voxel Exchange at all different numbers of GPUs, and the Gradient Decomposition has a memory footprint of 0.18 Gigabytes at 4158 GPUs, while the Halo Voxel Exchange can scale to only 462 GPUs with a memory footprint of 0.48 Gigabytes per GPU.

\begin{figure}
\centering
\includegraphics[width=\linewidth]{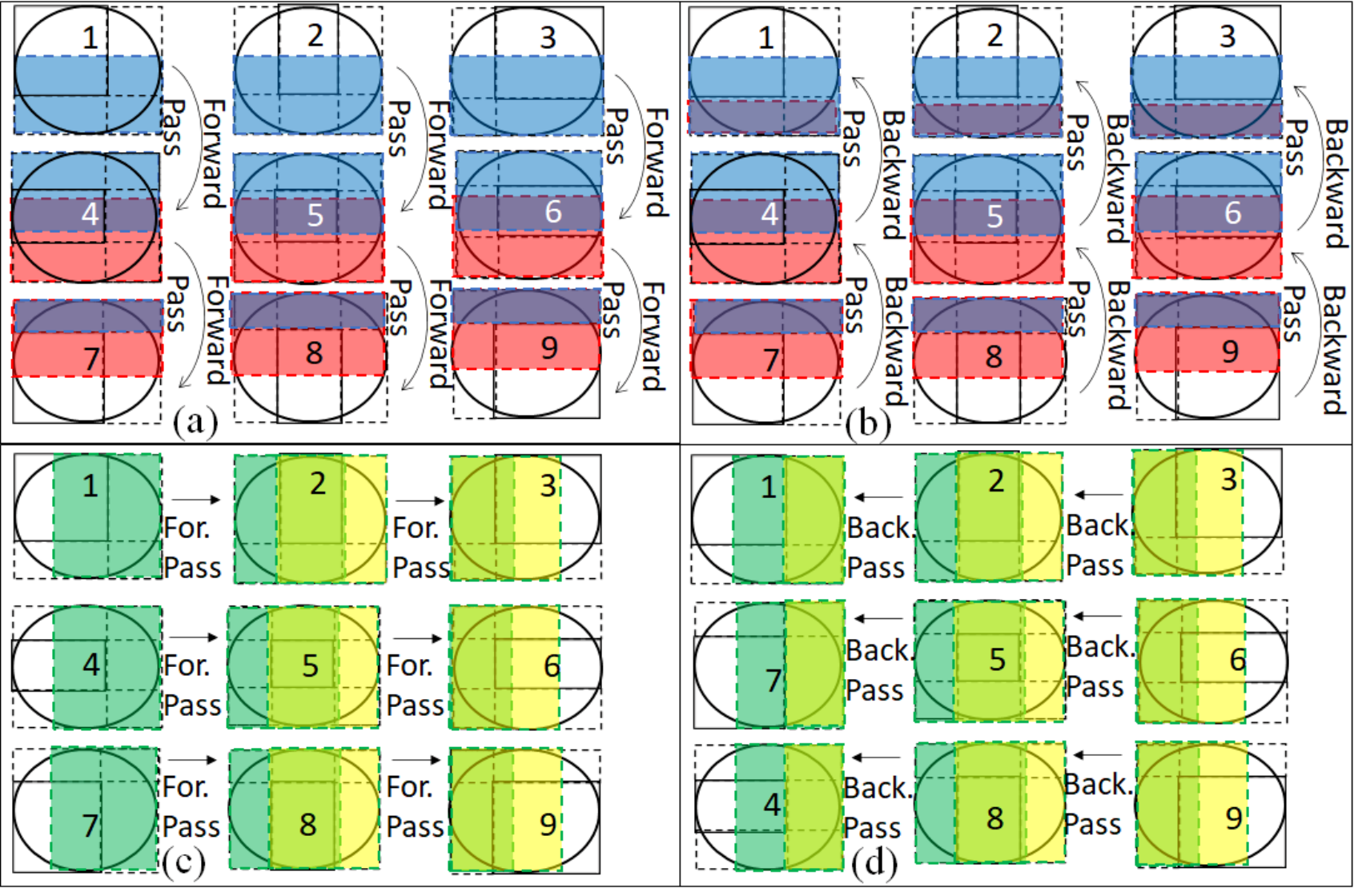}
 \caption{(a) Accumulated image gradients forward pass in the vertical direction. Notice that the bottom tiles receive accumulated gradients from the top tiles and the center row tiles, but the top tiles do not receive any gradients. (b) Accumulated image gradients backward pass in the vertical direction and the top tiles receives accumulated gradients from center row and bottom tiles. (c) Forward pass in the horizontal direction. (d) Backward pass in the horizontal direction.
   \label{fig:forward_backward} 
    } 
\end{figure}

\begin{figure*}
\centering
\includegraphics[width=.75\linewidth]{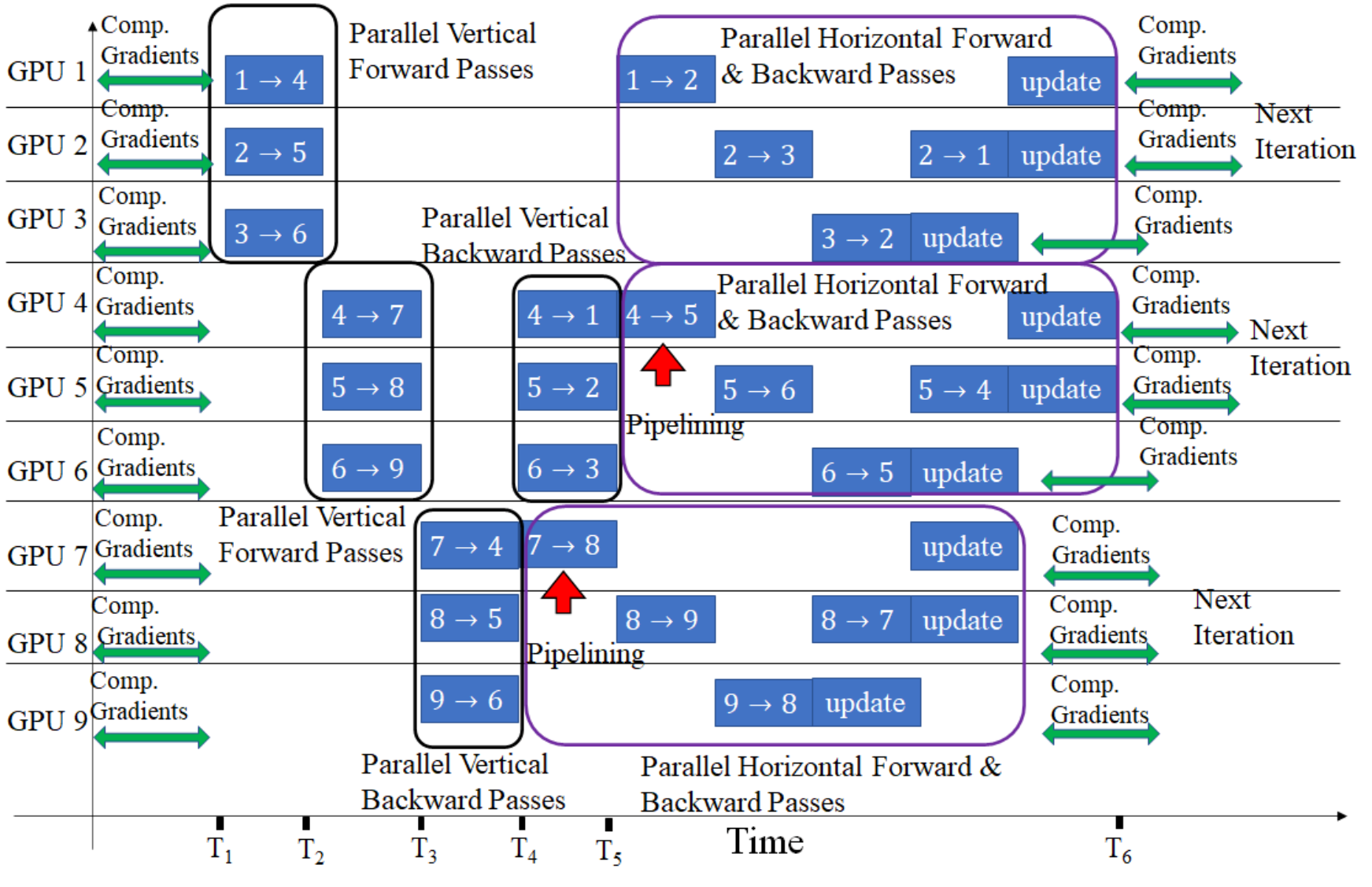}
 \caption{Asynchronous pipelines for parallel forward and backward passes for the example in Fig.~\ref{fig:forward_backward}.
 The 9 GPUs are in a logical 3x3 mesh and each is assigned a probe location circle.
 The horizontal axis is execution time sequence, and the vertical axis is GPUs. Green arrows represent the gradients compute time for each tile. Red arrows point to where pipelining happens. Notice that the forward and backward passes in both vertical and horizontal directions are executed in parallel. 
   \label{fig:pipeline} 
    } 
\end{figure*}

Another important distinction between the two algorithms is that the Gradient Decomposition method distributes image gradients into tiles and add image gradients to each other. In contrast, the Halo Voxel Exchange distributes and copy-pastes image voxels among tiles. 
By accumulating image gradients in the overlap regions among neighboring tiles, the Gradient Decomposition method smooths the transition from a tile to its neighboring tile in the lower spatial frequencies, thereby removing possible seam artifacts near the border of each tile and lead to superior image quality. In contrast, the Halo Voxel Exchange method directly copies and pastes the voxels among tiles, and thereby is sensitive to any minor voxel intensity differences among neighboring tiles and can easily cause artificial seam artifacts.
An example image artifacts comparison is provided in Fig.~\ref{fig:artifact} in the experiment section, showing that the Gradient Decomposition method can eliminate all image seam artifacts that are clearly visible in the Halo Voxel Exchange method.

\section{Forward and Backward Accumulated Gradients Pass}
\label{sec:forward_backward}

A special case that the direct neighbor gradient accumulations, discussed in Sec.~\ref{sec:image_gradient}, cannot handle is when the probe overlap ratio is high enough that each probe location circle not only overlaps with its direct neighbors, but also overlaps with indirect neighbors. For instance, we reuse the high overlap ratio example from Fig.~\ref{fig:halo_voxel_exch1}(f) and show its decomposed gradient tiles and halos in Fig.~\ref{fig:gradient}(c). If we only look at tiles 1, 4 and 7 and perform gradient accumulations among direct neighboring tiles as discussed in Sec.~\ref{sec:image_gradient}, tiles 1 and 4 add image gradients to each other in the blue overlap region, while at the same time tiles 4 and 7 add image gradients to each other in the red overlap region. After the gradient accumulations, some tiles, such as tile 4, receive all image gradients for its overlap region, but some other tiles, such as tiles 1 and 7, never receive each other's image gradients, despite that the yellow circle at tile 1 overlaps with both the red circle at tile 4 and the green circle at tile 7.

To address the above issue among non-adjacent tiles, we use two directional forward and backward passes to accumulate image gradients and Fig.~\ref{fig:forward_backward} explains the operations. Before the explanation, let us first introduce the concept of tile rows and tile columns here. A tile row is a group of tiles along the same row and tiles 1, 2, 3 are a tile row, and tiles 4, 5, 6 are in a different tile row. A tile column is a group of tiles along the same column and tiles 1, 4, 7 are a tile column, and 2, 5, 8 are a different tile column. 

At the start of forward and backward passes, each GPU creates an individual accumulated gradient buffer and stores a copy of its image gradients in the buffer. For the vertical direction forward pass in Fig.~\ref{fig:forward_backward}(a), the GPUs at the top tiles (1, 2 and 3) use point-to-point communications to add its accumulated gradient buffers to the buffers for the GPUs at the tiles below (tiles 4, 5, and 6) in the blue overlap region, with tile 1 added to tile 4 in the same tile column, tile 2 added to tile 5 and tile 3 added to tile 6. Then, the GPUs at the center tile row (4, 5, and 6) add their accumulated buffers to the buffers for the GPUs at the bottom tiles 7, 8 and 9, in the red overlap region. After the vertical direction forward pass, notice that the accumulated buffers for the bottom tiles 7, 8 and 9 are not only added with gradients from the adjacent center tile row in the red overlap region, but are also added with gradients from the top tiles in the overlap between the top and the bottom tiles.

For the vertical direction backward pass from the bottom tiles back to the top tiles in Fig.~\ref{fig:forward_backward}(b), the bottom tiles' accumulated gradient buffers are sent back to GPUs at the center tile row (4, 5, 6). Then, the accumulated buffers from the bottom tiles \textbf{replaces} the gradient buffers for the center tile row in their overlap region, so that the center row and the bottom tiles have the same gradients in the overlap between the two. Finally, GPUs at the center tile row (4, 5, and 6) sends their accumulated gradient buffers back to the top tiles and replaces the top tiles' buffers with the gradients from the buffers at the center tile row in their overlap region. After the vertical direction backward pass, notice that the top tiles have the accumulated gradients from its own gradients, the overlap region gradients from the adjacent center tile row, and the overlap region gradients from the bottom tiles. 

In addition to vertical direction forward and backward passes, the Gradient Decomposition also needs to perform horizontal direction forward and backward passes to accumulate gradients from the leftmost to the rightmost tiles and vice versa. For the horizontal direction forward pass in Fig.~\ref{fig:forward_backward}(c), the gradients in the accumulated buffers from the leftmost tiles are added to the buffers for the center tile column (tiles 2, 5, and 8), which in turn are then added to the buffers for the rightmost tiles. The backward pass in Fig.~\ref{fig:forward_backward}(d) replaces the gradient buffers for the center tile column with the buffers for the rightmost tiles in their overlap, and then does the same operation between the leftmost tiles and the center tile column.  Finally, each GPU uses their gradients in the buffers to independently update their tiles and the algorithm repeats itself in iterations until converged.

\begin{algorithm}[t]
\caption{Asynchronous Pipelining for Parallel Passes} 
\label{alg:appp}
\begin{algorithmic}[1]

\INPUT Diffraction measurements $y$.
\OUTPUT The final reconstruction $V$.

\For{each GPU $k$ from 1 to $K$ run \textbf{in parallel}}
\State Create accumulated gradient buffer $\text{AccBuf}_k$.
\State Receive decomposed extended tile $V_k$.
\While{ not converged}
\For{each probe location $i$ for the $k^{\it th}$ GPU:} \label{step:start-probe-iter}
\State Compute individual gradient for tile $V_k$: \vfill $\frac{\partial f_i}{\partial V_k} \gets \frac{\partial \left( | y_i| - |G(p_i,V_k)|   \right)^2}{\partial V_k}$
\State $\text{AccBuf}_k \pluseq  \frac{\partial f_i}{\partial V_k} $ \label{step:buffer-tracking}  
\State $V_k \gets V_k - \alpha \frac{\partial f_i}{\partial V_k}$
\If{$i \mod T ==0$} \label{step:if-mod}
\State Vertical forward pass on $\text{AccBuf}_k$ 
\State Vertical backward pass on $\text{AccBuf}_k$
\State horizontal forward pass on $\text{AccBuf}_k$
\State horizontal backward pass on $\text{AccBuf}_k$
\State $\frac{\partial f_i}{\partial V_k} \gets \text{AccBuf}_k $ 
\State $V_k \gets V_k - \alpha \frac{\partial f_i}{\partial V_k}$ \label{step:update_tiles_again}
\State $\text{AccBuf}_k \gets 0$
\EndIf
\EndFor \label{step:end-probe-iter}
\EndWhile
\State Abandon halos and stitch together non-halo tiles into a final reconstruction $V$
\EndFor
\end{algorithmic}
\end{algorithm}

\section{Asynchronous Pipelines for Parallel Passes}
\label{sec:async_pipelines}
The forward and backward passes create dependencies between them. The input to the horizontal direction backward pass in Fig.~\ref{fig:forward_backward}(b) depends on the  output of the forward pass in the same direction in Fig.~\ref{fig:forward_backward}(a). The same dependency applies to vertical direction forward and backward passes. In addition, there is a cross-direction dependency between the input to the horizontal direction forward pass in Fig.~\ref{fig:forward_backward}(c) and the output to the vertical direction backward pass in Fig.~\ref{fig:forward_backward}(b).

A natural choice to address the dependencies is global all-reduce operations. However, all-reduce has large communication overhead and significantly decreases scalability. To scale to many GPUs, we introduce the Asynchronous Pipelining for Parallel Passes (APPP) technique. Fig.~\ref{fig:pipeline} describes its operations and reuses the 9 tile example from Fig.~\ref{fig:forward_backward}. The horizontal axis of the figure represents time sequence, and the vertical axis represents GPUs 1 to 9. From time points 0 to $T_1$, each GPU independently computes their gradients from their measurements. Then from time $T_1$ to $T_2$, vertical direction forward passes are performed in parallel across different tile columns. The accumulated gradient buffers are added from the GPU at tile 1 to the GPU at tile 4, while at the same time tile 2 is added to 5 and tile 3 is added to 6. To help reduce communication time, we use MPI asynchronous non-blocking operations (isend and irecv) operations to send and receive accumulated gradient buffers across GPUs and therefore some communication cost can be hidden in the gradient computations from time 0 to $T_1$. Similarly, the vertical direction forward and backward passes are performed in parallel from time $T_2$ to $T_3$ and time $T_3$ to $T_5$, respectively.

Notice that GPU 7 does not wait for the backward pass to complete. While GPUs 4, 5, 6 are performing backward passes from $T_4$ to $T_5$, GPU 7 is simultaneously performing horizontal direction forward pass with other GPUs in the same tile rows (tiles 7, 8 and 9), thereby achieving cross-direction pipelining between the vertical direction backward pass and the horizontal direction forward pass. Similarly, GPU 4 does not wait for backward pass to complete and performs cross-direction pipelining immediately after it sends its accumulated buffer to GPU 1. After pipelining, GPUs across different tile rows, where each tile row is highlighted by purple rectangles in Fig.~\ref{fig:pipeline}, perform horizontal direction forward and backward passes in parallel. At the end, each tile is updated based on the gradients in the buffer. Then the algorithm repeats itself in iterations until converged.

Despite of its advantage for reducing communication overhead, the asynchronous parallel pipelines can slightly worsen GPU waiting time. In the example of Fig.~\ref{fig:pipeline}, GPU 9 is the first to finish updates and enter the next iteration, while GPUs 1, 2, 4, and 5 are the last to finish. Therefore, GPUs might wait on each other to finish their gradient computations before performing parallel passes. The benefit from reducing communication overhead, however, outweighs the loss from increased GPU waiting time, especially when the number of GPUs are many. With many GPUs and few computations for each, GPUs are less likely to wait on each other's gradient computations. In addition, the MPI asynchronous non-blocking operations can further hide the GPU waiting time. Our large-scale material image reconstruction in Fig.~\ref{fig:runtime_breakdown} of the experiment section demonstrates that the GPU communication overhead for 77 nodes (462 GPUs) is 16 times smaller with the APPP than without the APPP. Meanwhile, the GPU waiting time decreases from 263 minutes for 24 GPUs to about a second for 462 GPUs.

To further reduce communication overhead, we use a delayed gradient accumulation technique, explained in Alg.~\ref{alg:appp}, to lower the communication frequency. In the algorithm, we use parameter $T$ to represent how many times in an iteration we want to communicate across GPUs through forward and backward passes. When $T>1$, the communications are less frequent and the overhead is smaller than performing passes after every probe location.
From steps~\ref{step:start-probe-iter} to~\ref{step:end-probe-iter} of Alg.~\ref{alg:appp}, each GPU computes its individual gradients and updates its tile after going through each probe location without performing any forward or backward pass. At the same time, the accumulated gradient buffer keep track of the gradient changes by adding gradients at each probe location to the buffer in step~\ref{step:buffer-tracking}. Then after every $T$ number of probe locations in step~\ref{step:if-mod}, the gradients are accumulated in the buffers through the asynchronous parallel pipelines. The tiles are then updated again in step~\ref{step:update_tiles_again} using the accumulated gradients and the buffer is reset to zeros. Alg.~\ref{alg:appp} then repeats itself until reaching convergence. Finally, the non-halo tiles are stitched together for a final reconstruction.

\section{Results}
\label{sec:experiments}
\subsection{Datasets, Computing Platform, Parameters and Software}
\label{subsec:dataset_platform}

\begin{figure}
\centering
   \includegraphics[width=.6\linewidth]{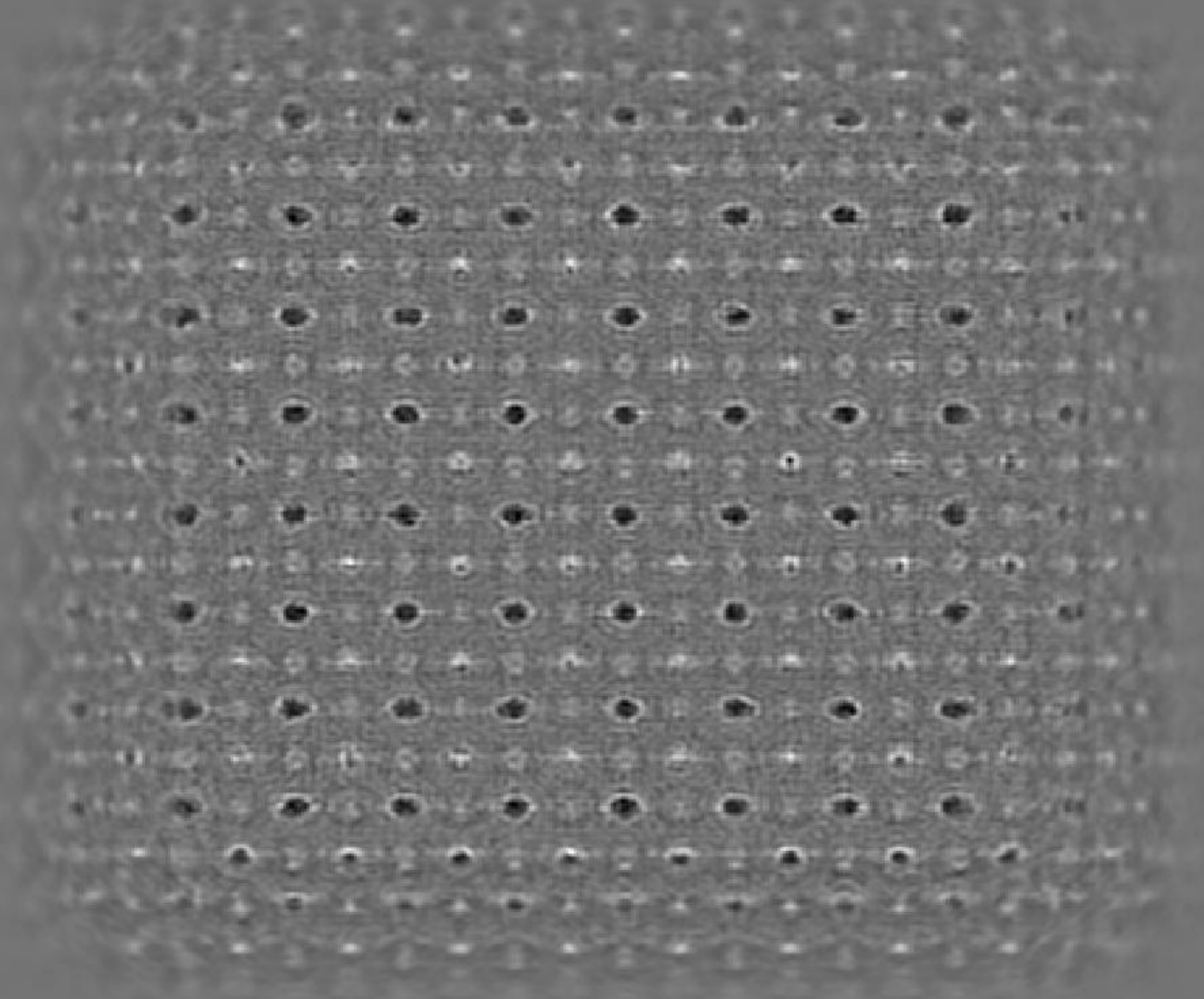}
 \caption{ An example image for Lead Titanate material and each circle in the image represents a small group of atoms.
   \label{fig:example_slice} 
    } 
\end{figure}

\begin{table}
\begin{tabular}{ |c|c|c| } 
 \hline
 Sample Name &  Lead Titanate small & Lead Titanate large \\ 
Measurements $y$ size & $1024 \times 1024 \times 4158$ & $1024 \times 1024 \times 16632$ \\ 
Reconstruction $V$ size  & $1536 \times 1536 \times 100$  & $3072 \times 3072 \times 100$ \\
Image resolution & $10 \times 10 \times 125$ ${pm}^3$ & $10 \times 10 \times 125$ ${pm}^3$ \\
 \hline

\end{tabular}
\caption{\label{table:datasets-size}Dataset sizes for measurements and reconstructions.}

\end{table}

\textbf{Datasets:} Lead Titanate
(PbTiO\textsubscript{3}) is an important material under research for manufacturing ultrasound transducers and sensors as well as high-value ceramic capacitors for its superior material property. An example image slice for Lead Titanate is provided in Fig.~\ref{fig:example_slice}. In this section, we evaluate all our performance numbers on two Lead Titanate
simulated material datasets, imaged with electron microscope at 25 nanometer defocus, 200 kiloelectron volts and a probe forming aperture of 30 milliradians. Table~\ref{table:datasets-size} summarizes the sizes of the two datasets, and one dataset has 4158 probe locations and the other dataset has 16632 probe locations.   At each probe location, the diffraction measurements size is $1024^2$. Therefore, measurements $y$ for the small Lead Titanate material dataset with 4158 probe locations has a total size of $1024 \times 1024 \times 4158$, and the measurements size for the large dataset with 16632 probe locations has a total size of $1024 \times 1024 \times 16632$. As to the reconstruction $V$ size, the small dataset reconstruction size is $1536 \times 1536 \times 100$, where $1536^2$ is the size in the x-y plane and 100 is the number of slices, and each voxel size is $10 \times 10 \times 125$ Cubic Picometers. The large dataset reconstruction size is $3072 \times 3072 \times 100$ at the same image resolution.

\begin{table*}
\begin{subtable}{.52\linewidth}\centering
{
{\small
\begin{tabular}{|c|c|c|c|c|c|c|}
\hline
Nodes & 1  & 4  & 9 & 21 & 33 & 77 \\
\hline
GPUs & 6  & 24  & 54 & 126 & 198 & 462 \\
\hline
 \begin{tabular}{@{}c@{}}Memory footprint \\ per GPU (GB)\end{tabular} & 2.53  & 1.20 & 0.58 & 0.39 & 0.31 & 0.23 \\
\hline
Runtime (mins) & 360.0  & 73.0  & 20.6 & 11.5 & 5.5 & 3.0
 \\
\hline
\begin{tabular}{@{}c@{}}Strong scaling \\ efficiency\end{tabular} & $100\%$  & $123\%$ & $194\%$ & $149\%$ & $198\%$ & $158\%$
 \\
\hline
\end{tabular}
}
}
\caption{Gradient Decomposition for The Small Lead Titanate Dataset}
\end{subtable}
\begin{subtable}{.47\linewidth}\centering
{
{\small
\begin{tabular}{|c|c|c|c|c|}
\hline
Nodes & 1  & 4  & 9 & 21 \\
\hline
GPUs & 6  & 24  & 54 & 126 \\
\hline
 \begin{tabular}{@{}c@{}}Memory footprint \\ per GPU (GB)\end{tabular} & 2.80  & 1.20 & 0.78 & NA  \\
\hline
Runtime (mins) & 463.3  & 95.3  & 43.7 & NA 
 \\
\hline
\begin{tabular}{@{}c@{}}Strong scaling \\ efficiency\end{tabular} & $100\%$  & $121\%$ & $118\%$ & NA 
 \\
\hline
\end{tabular}
}
}
\caption{Halo Voxel Exchange for The Same Dataset}
\end{subtable}

\caption{Performance comparison for the proposed Gradient Decomposition and the state-of-the-art Halo Voxel Exchange methods on the small Lead Titanate dataset.}

\label{table:performance-small}
\end{table*}

\begin{table*}
\begin{subtable}{.52\linewidth}\centering
{
{\small
\begin{tabular}{|c|c|c|c|c|c|c|}
\hline
Nodes & 1  & 9  & 33 & 77 & 154 & 693 \\
\hline
GPUs & 6  & 54  & 198 & 462 & 924 & 4158 \\
\hline
 \begin{tabular}{@{}c@{}}Memory\\ (GB)\end{tabular} & 9.14  & 1.54 & 0.66 & 0.42 & 0.32 & 0.18 \\
\hline
Runtime (mins) & 5543.0 & 183.0 & 37.5 & 14.2 & 7.0 & 2.2
 \\
\hline
\begin{tabular}{@{}c@{}}Strong scaling \\ efficiency\end{tabular} & $100\%$  & $336\%$ & $448\%$ & $509\%$ & $518\%$ & $364\%$
 \\
\hline
\end{tabular}
}
}
\caption{Gradient Decomposition for The Large Lead Titanate Dataset}
\end{subtable}
\begin{subtable}{.47\linewidth}\centering
{
{\small
\begin{tabular}{|c|c|c|c|c|}
\hline
Nodes & 1  & 9  & 33 & 77 \\
\hline
GPUs & 6  & 54  & 198 & 462 \\
\hline
 \begin{tabular}{@{}c@{}}Memory\\ (GB)\end{tabular} & 9.47 & 1.8 & 0.78 & 0.48  \\
\hline
Runtime (mins) & 7213.3 & 271.7 & 59.2 & 189.5
 \\
\hline
\begin{tabular}{@{}c@{}}Strong scaling \\ efficiency\end{tabular} & $100\%$  & $295\%$ & $369\%$ & $49\%$ 
 \\
\hline
\end{tabular}
}
}
\caption{Halo Voxel Exchange for The Same Dataset}
\end{subtable}

\caption{Performance comparison for the two methods on the large Lead Titanate dataset.}

\label{table:performance-large}
\end{table*}

\begin{figure*}
\begin{subfigure}{.57\linewidth}
\centering
\includegraphics[width=\linewidth]{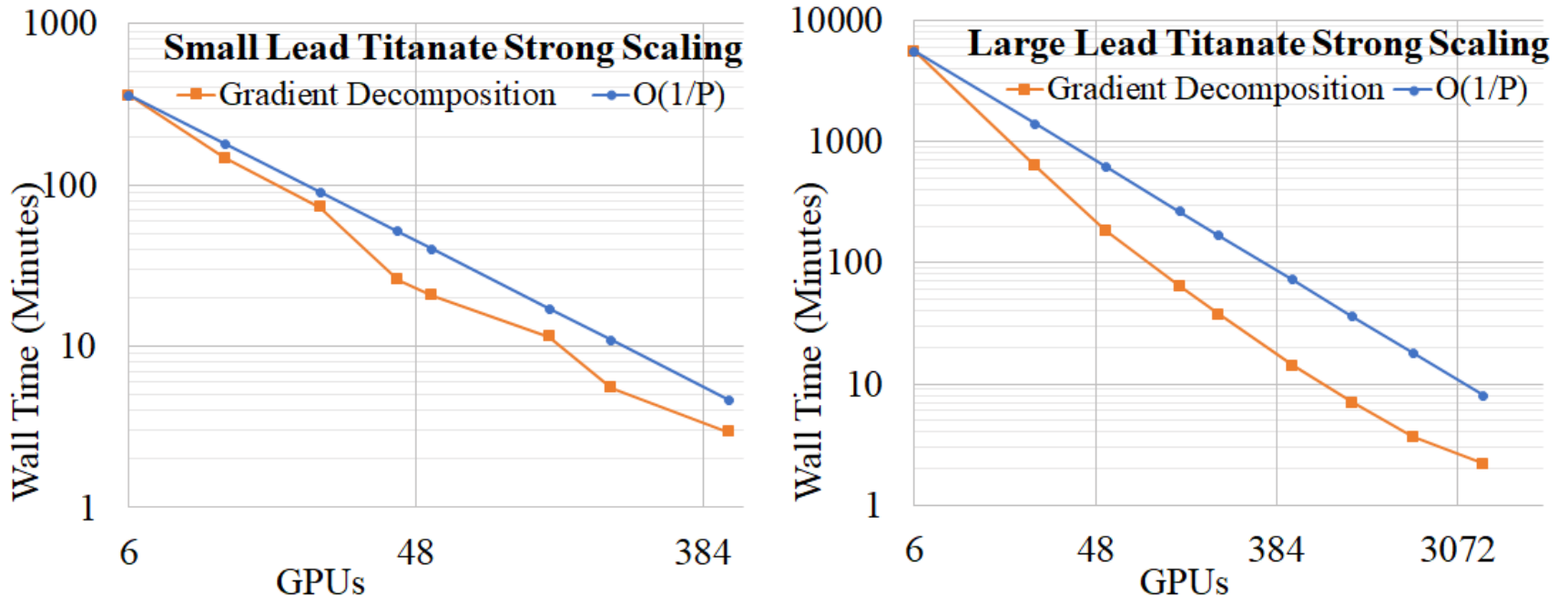}
\caption{
    } 
\label{fig:scaling} 
\end{subfigure}
\begin{subfigure}{.43\linewidth}
\centering
\includegraphics[width=\linewidth]{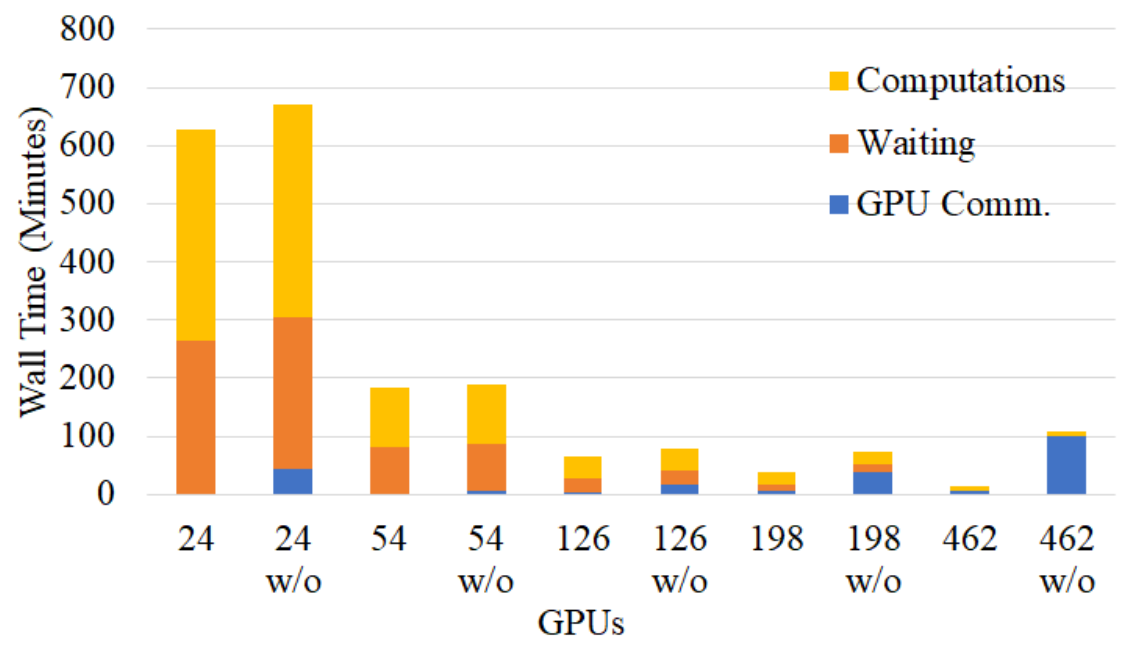}
\caption{
    } 
\label{fig:runtime_breakdown} 
\end{subfigure}
\caption{(a) Strong scaling for both datasets. (b) Runtime breakdown for the large Lead Titanate datasets, with and without the Asynchronous Pipelines for Parallel Passes (APPP). ``w/o" represents runtimes without the APPP.}
\end{figure*}

\textbf{Computing platform:} 
The computing platform is the Oak Ridge National Lab Summit supercomputer with 6 NVIDIA Volta V100 GPUs and 2 POWER9 CPUs per node.
Nodes are connected via Mellanox EDR 100G InfiniBand Non-blocking Fat Tree network.
Each POWER9 CPU in the node is densely-connected to 3 GPUs with Nvidia
NVlinks, where each link has 50 GB/s one-way and 100 GB/s bidirectional bandwidth, and the two CPUs for each node are connected via an X bus with 64 GB/s bidirectional
bandwidth. Each CPU has 22 cores (4 hardware threads for each) and 256 GB DRAM memory. Each GPU has 80 streaming multiprocessors and 16 GB memory. More information on the hardware can be found in the artifact description appendix. 

\textbf{Reconstruction parameters:} The Gradient Decomposition method uses a halo width of 600 Picometers for each tile. The Halo Voxel Exchange uses two extra rows of probe locations for each tile and has a halo width of 890 Picometers to cover all probe locations. Unless otherwise noted, all runs for the Gradient Decomposition method perform bi-directional forward and backward passes once per iteration, and an iteration is defined as a cycle through all the probe locations.

\textbf{Software}: The image gradient decomposition software for ptychographic reconstruction is available at \url{https://code.ornl.gov/ai-ptychography/ptychopath/-/tree/23-basic-ddp-example}

\subsection{Reconstruction Runtimes and Memory Footprint}
\label{subsec:recon_memory}
Tables~\ref{table:performance-small} and~\ref{table:performance-large} summarize the performance comparison between the Gradient Decomposition and Halo Voxel Exchange methods for both Lead Titanate datasets. The first two rows show the number of nodes with 6 GPUs for each node. The third row shows the average peak memory footprint in Gigabytes (GBs) per GPU. We can notice that both algorithms reduce memory footprint with increased number of GPUs, although the Gradient Decomposition method achieves more memory reduction and reduces memory footprint for the large dataset from more than 9 GBs per GPU to less than 0.18 GB per GPU by scaling from 6 to 4158 GPUs. In comparison, the smallest memory footprint for the Halo Voxel Exchange method is at 0.48 GB, which is three times larger than that for the Gradient Decomposition. The Halo Voxel Exchange method has worse memory reduction capability than the Gradient Decomposition for two reasons. First, the Halo Voxel Exchange has poorer scalability and thereby is less capable to distribute the data storage across GPUs. Second, the Halo Voxel Exchange requires extra memory per GPU for the storage and computations for the additional probe locations in its halos, explained before in Figs.~\ref{fig:halo_voxel_exch1}(d)-(e).

The fourth row of the tables shows the average runtimes in minutes for fixed 100 iterations so that convergence speed is not a factor in the reported runtimes.
For the large Lead Titanate dataset, the Gradient Decomposition algorithm completes reconstruction in 2.2
minutes by scaling to 4158 GPUs.
To our knowledge, this is the fastest reconstruction in near real time for an electron ptychography dataset of this size.
In comparison, the Halo Voxel Exchange fastest runtime is at 59.2 minutes by scaling to 198 GPUs. If scaling to more than 198 GPUs, the Halo Voxel Exchange has a sharp increase in runtime due to increased communication cost and computation burden for the additional probe locations. We also want to point out that the Halo Voxel Exchange has an inherent algorithmic limitation for its scalability. Since each tile pastes its voxels to the neighboring tile's halo through memory copy operations, each tile needs to be sufficiently large to occupy the neighboring tile's halos. Otherwise, the neighboring tiles cannot be consistent with each other. Unfortunately, more GPUs lead to a smaller tile for each GPU. Therefore, the tile size constraint limits the total number of GPUs available for scaling. In Table~\ref{table:performance-small}(b), we use ``NA" to indicate that The Halo Voxel Exchange cannot scale to more than 54 GPUs for the small Lead Titanate dataset due to the tile size constraint. 

\subsection{Strong Scaling Efficiency}
\label{subsec:strong_scaling}
The fifth rows of the tables show the strong scaling efficiency compared to 1 node performance. The complete strong scaling data points for the Gradient Decomposition can be found in Fig.~\ref{fig:scaling}. In addition, Fig.~\ref{fig:scaling} also provides a comparison with the ideal linear speedup runtimes and uses notation $O(1/P)$ to indicate the linear performance, where $P$ is the number of GPUs.  
We can observe that both algorithms achieve super-linear speedups from the tables and Fig.~\ref{fig:scaling}. The super-linear speedups happen due to (1) reduced computations, (2) improved cache hits when the number of GPUs increase, and (3) decreased GPU waiting time with more GPUs, which will be explained later in Sec.~\ref{subsec:communication}.

To compute Eqn.~(\ref{eqn:math_formulation}), function $G$ computes Fourier Transform and inverse Fourier Transform for each image slice in sequence as explained before in Sec.~\ref{sec:math_formulation} and the computation cost for function $G$ increases non-linearly with an $N\log N$ growth order when the problem size increases. Therefore when we increase the number of GPUs, the number of computations for function $G$ for each GPU decreases non-linearly, leading to a super-linear speedups in Fig.~\ref{fig:scaling}. In addition, since each GPU has a smaller problem size and fewer data to operate on when the number of GPU increases, the data for each GPU are more likely to fit into GPU caches, leading to improved cache hits, faster Fourier Transform and super-linear speedups. To corroborate our finding, we use the NVIDIA Nsight Compute to profile the GPU cache hit rate for computing function $G$. From 24 GPUs to 54 GPUs, for example, The Nsight Compute show that the L1 cache hit rate for the $G$ function computations in the Gradient Decomposition increases from $44 \%$ to $59 \%$, and the total memory throughput increases from $34\%$ to $49\%$.

\subsection{GPU Communication and Waiting Overhead}
\label{subsec:communication}
Besides improved cache hits, another reason that contributes to super-linear speedup is that the Gradient Decomposition method has a decreased GPU waiting time when the number of GPUs increases while maintaining a small communication overhead for gradient accumulations. 
Fig.~\ref{fig:runtime_breakdown} is the runtime breakdown for the Gradient Decomposition algorithm on the large Lead Titanate dataset from 24 to 462 GPUs. In addition, the figure shows the runtime breakdown comparison for both with and without the Asynchronous Pipelining for Parallel Passes (APPP) technique discussed in Sec.~\ref{sec:async_pipelines}. In Fig.~\ref{fig:runtime_breakdown}, the yellow bars represent the computation times, orange bars represent the GPU waiting time, and the blue bars represent the GPU communication overhead for gradient accumulations by performing asynchronous MPI send and receive operations. The keyword ``w/o" in the figure indicates that the corresponding runtimes do not use the APPP technique, while runtimes without the keyword use the APPP. 

One observation from the figure is that with the asynchronous parallel pipelines, the communication overhead for the Gradient Decomposition remains low even at 462 GPUs. In comparison, the runtimes are dominated by communication overhead at 462 GPUs when the APPP is turned off. 
Another observation is that the GPU waiting time is long when GPUs are few, but the waiting time steadily decreases with increased number of GPUs. When the GPUs are few, each GPU has a large tile and many gradient computations. Therefore, there is a high probability for the GPUs to wait for each other's gradient computations before performing gradient accumulations asynchronous pipelines. When the number of GPUs increases, however, each GPU has fewer computations and thereby has a smaller probability to wait for each other's gradient computations, leading to decreased waiting time.

\subsection{Image Artifacts}
\label{subsec:artifacts}
\begin{figure}
\centering
\includegraphics[width=.6\linewidth]{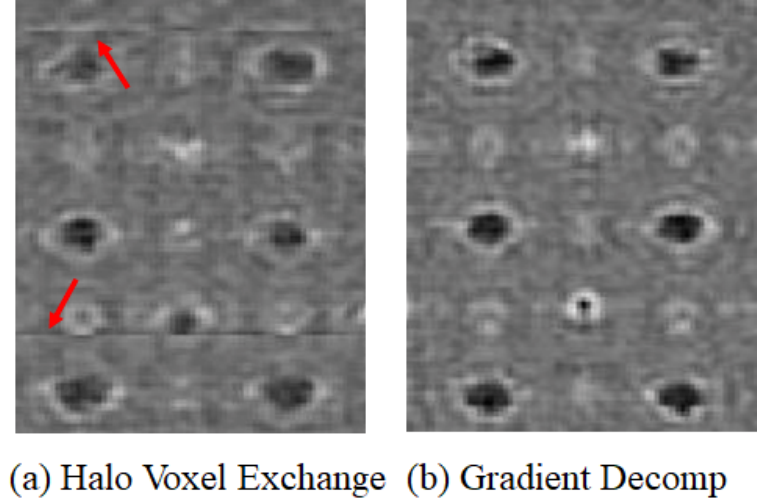}
\caption{(a) Seam artifacts from the Halo Voxel Exchange. (b) Artifacts eliminated in the Gradient Decomposition.
} 
\label{fig:artifact} 
\end{figure}

An important advantage for the Gradient Decomposition method is that it can eliminate image seam artifacts near the border of tiles, while the Halo Voxel Exchange method cannot. Fig.~\ref{fig:artifact} shows an example region in the reconstruction where the Halo Voxel Exchange in Fig.~\ref{fig:artifact}(a) has the artificial border artifacts, pointed by red arrows in the figure, while the Gradient Decomposition in Fig.~\ref{fig:artifact}(b) does not. 
By accumulating image gradients in the overlap regions, the Gradient Decomposition method smooths the transition from a tile to its neighboring tile in the lower spatial frequencies, thereby removing  seam artifacts near the border of each tile and lead to superior image quality. In contrast, the Halo Voxel Exchange method directly pastes voxels to each other through memory copy operations. Thereby the Halo Voxel Exchange is sensitive to any minor voxel intensity differences among neighboring tiles and can easily cause artificial seam artifacts.

\subsection{Convergence}
\label{subsec:convergence}
\begin{figure}
\centering
\includegraphics[width=\linewidth]{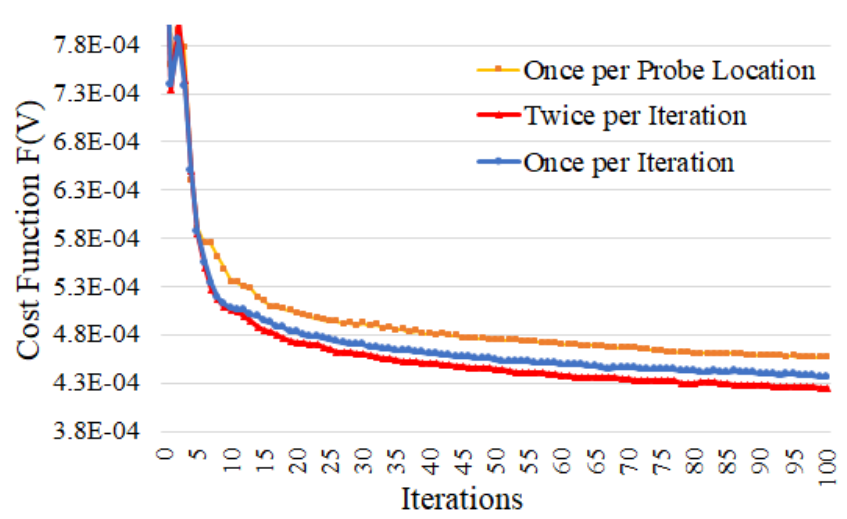}
\caption{ Convergence rate for the Gradient Decomposition with three different communication frequency for parallel passes.
} 
\label{fig:convergence} 
\end{figure}

The asynchronous parallel pipelines use parameter $T$ in Alg.~\ref{alg:appp} to delay gradients exchanges and reduce communication frequency. Despite its clear advantage for lowering communication overhead, it is important to understand how significant frequency reduction can impact the algorithmic convergence. Fig.~\ref{fig:convergence} shows the convergence rate for the Gradient Decomposition method on 42 GPUs with three different communication frequencies, and the vertical axis is the value of the cost function $F(V)$ in Eqn.~(\ref{eqn:math_formulation}). The yellow curve is the convergence rate for $T=1$ that performs parallel passes at every probe location. The red curve is the convergence when performing the parallel passes twice per iteration, and the blue curve is the convergence for once per iteration. Since the probe locations overlap with each other, high frequency of parallel passes can in fact cause convergence overshooting for the probe overlap regions. Therefore, the red and blue curves (twice and once per iteration) not only have much smaller communication overhead with a reduced frequency than the orange curve (once per probe location), but also have slightly faster convergence rate.    

\section{Conclusion}
\label{sec:conclusion}
Ptychographic reconstruction is a widely-used 3D imaging technique that has a growing impact on the material, biology, and security imaging communities. The computation cost and the memory constraint are major bottlenecks for scientific advancements in these fields and hinder the desirable image resolution for ptychographic imaging. This paper describes the first gradient-based decomposition method for ptychographic imaging without introducing image artifacts. In addition, this paper introduces an asynchronous pipelining technique to achieve ultra-high-resolution imaging in near real-time by significantly improving computation speed, reducing memory footprint, and enabling massive parallelism.

The technical innovations of this paper will yield immediate
results to the imaging community. Our experiments demonstrate that the Gradient Decomposition method can reconstruct a large material dataset with 16632 probe location at an extremely high resolution of 10 picometer per voxel size, within 2.2 minutes on 4158 GPUs, and reduce its memory footprint from 9.14 GB per GPU to less than 0.18 GB. 
Therefore, material scientists can image their material samples at a high resolution without a prohibitive memory constraint, and can significantly reduce their wall-clock delay for the results of experiments.

\bibliographystyle{IEEEtran}
\bibliography{main}

\end{document}